\documentstyle[12pt]{article}
\begin{document}

\begin{titlepage}
\begin{flushright}
\par\vglue -2cm
nucl-th/9606007\\
ISN 96--66
\end{flushright}
\vfill
\begin{center}
{\Large{\bf NEUTRON--ANTINEUTRON OSCILLATIONS}}
\vskip .3cm
{\Large{\bf  AT THE SURFACE OF NUCLEI\footnote{Invited Talk by JMR at  
The
International Workshop on Baryon Instability Search in $p$-Decay and
$n-\bar n$ Oscillation Experiment, Oak Ridge, Tennessee, March  
28--30,
1996, to be published in the Proceedings.}   }}\\
\vskip .5cm
{\bf Carl B. Dover}\\
\vskip .1cm
{\small Physics Department}\\
{\small Brookhaven National Laboratory}\\
{\small Upton, N.Y., U.S.A.}\\
\vskip .25cm
{\bf Avraham Gal}\\
\vskip .1cm
{\small Racah Institute of Physics}\\
{\small The Hebrew University, Jerusalem, Israel}\\
\vskip .25cm
{\bf Jean-Marc Richard}\\
\vskip .1cm
{\small Institut des Sciences Nucl\'eaires--CNRS--IN2P3}\\
{\small Universit\'e Joseph Fourier, Grenoble, France}\\
\vglue 1.5cm
{\bf Abstract}\\
\vglue.2cm
\end{center}
We discuss some aspects of possible neutron--antineutron oscillations  
in
nuclei. The phenomenon  occurs mostly at the  surface of nuclei, and
hence {\sl i)} is not very sensitive to medium corrections and {\sl  
ii)}
makes use of the antinucleon-nucleus interaction in a region probed  
by
experiments at CERN.
\par
\vfill
\begin{flushleft}
ISN-96-66\\
nucl-th/9606007\\
\today.
\end{flushleft}
\end{titlepage}
The relevance of neutron--antineutron oscillations for testing  
physics
beyond the standard model was stressed in several contributions at  
this
Workshop, in particular by Mohapatra \cite{Mohapatra}.
The question now is how to detect neutron--antineutron oscillations,  
or how
to set an upper limit on their rate. We refer to Alberico's talk for  
a
comprehensive survey \cite{Alberico}. In the present contribution, we
wish to stress  some properties of the potential-model approach,
which in our opinion make it rather reliable.

Consider for instance an S-wave neutron in  deuterium. It is governed  
by a
radial Schr{\"o}dinger equation
\begin{equation}
\label{n-deut-S}
u''(r)+m\left[E-V_n(r)\right] u(r)=0,
\end{equation}
where notations are obvious. With a $n\leftrightarrow\bar{n}$  
transition
potential  $\epsilon$, the wave function gets an antineutron  
component
$\bar u(r)$ which to leading order is given by the inhomogeneous  
equation
\begin{equation}
\label{nbar-deut-S}
\bar u''(r)+m\left[E-V_{\bar n}(r)\right] \bar u(r)=m\epsilon u(r),
\end{equation}
where $E$ and $u$ are now frozen. This  leads to an estimate of the  
decay
width of deuterium in terms  of the annihilation part of the  
antineutron
potential
\begin{equation}
\label{width-deut-S}
-{\Gamma\over2}=-{1\over 2T}=
\int\limits_0^\infty\bar u^2(r){\rm Im}V_{\bar n}(r) {\rm d}r.
\end{equation}

It is rather straightforward to generalise these equations to larger  
nuclei,
once each neutron  is described by its  appropriate shell-model
wave-function.

Such a set of equations gives results which are stable with respect  
to
variations of the basic ingredients, as stressed, e.g., by Ericson  
and
Rosa-Clot \cite{Ericson-RC} in a different context. Moreover, this  
formalism
provides the relative contribution of each neutron shell, and within  
a
shell, the weight of the various parts in the integration over the  
distance
$r$.

This analysis was carried out in Ref.\ \cite{DGR1}, and its results  
were
confirmed in further investigations \cite{DGR2}: neutron--antineutron
oscillations, if any, occur mostly in outer shells, and near the  
surface of
the nucleus. In particular:

{\large $\bullet$} nuclei with weakly-bound neutrons offer more  
favourable
rates. As discussed during the oral presentation, heavy water would  
be
slightly better than ordinary water, because of the loosely-attached  
neutron
in deuterium.

{\large $\bullet$} the antineutron wave function is peaked outside  
the nuclear
density.

These properties survive changes in the assumed shape of the
neutron potential, and can be explicitly seen in toy models such as
square-well or separable potentials, for which calculations can be  
performed
analytically.

This peripheral character of nuclear instability can be explained as  
follows.
In  nuclear medium,  a neutron which considers the possibility of  
oscillating
is immediately refrained to do so when it feels the large gap between  
the
average potentials $V_n$ and  $V_{\bar n}$. Far away, this gap
vanishes, for both potentials go to zero. On the other hand, one  
would
not care too much about an antineutron at large distance from the  
centre, as
it has no nucleon to interact with. The best compromise takes place  
at the
surface, where the neutron is free enough to oscillate, and the  
medium is
still dense enough to annihilate freshly-produced antineutrons.

The peripheral character of neutron oscillations in nuclei has two  
important
consequences:

{\large$\bullet$} Contrary to the neutron potential $V_n$ (and its  
by-product
$u(r)$, the neutron wave function) which is well constrained by  
decades of
phenomenological studies, the antineutron potential $V_{\bar n}(r)$  
is not
well known inside the nucleus. However, experiments at the LEAR  
facility
of CERN with antiprotonic atoms, or with antiproton beams scattered  
on nuclear
targets, provide stringent constraints on this interaction near the  
nuclear
surface \cite{Dover}. This is precisely the domain we need for  
neutron
oscillations.

{\large$\bullet$} Deeply inside a nucleus, the transition operator  
$\epsilon$
could be renormalised, or receive new contribution. This is  
discussed, e.g.,
by Kabir \cite{Kabir}. The quasi-free neutrons which contribute to  
most of the
decay of the nucleus experience an $\epsilon$ potential which is  
identical or
very close to $\tau^{-1}$, where $\tau$ is the oscillation period of  
free
neutrons.

The comparison between free-neutron  and bound-neutron experiment has  
been
the subject of intense debates, and one of the authors (J.M.R.)   
would
like to acknowledge discussions on the subject at I.L.L., Grenoble,  
with the
late R.\ Marshak and W.\ Mampe. As long as negative results are  
obtained,
bound nuclei provide a rather reliable lower bound on the oscillation  
period
$\tau$. The scaling properties of  
Eqs(\ref{nbar-deut-S}-\ref{width-deut-S})
implies for $\tau$ a relation
\begin{equation}
\label{T-to-tau}
T=T_R\tau^2
\end{equation}
to the lifetime $T$ of a given nucleus. When the nuclear factor $T_R$  
is
computed with the typical medium-size nuclei used in proton-decay
experiments, one gets that $T\ge10^{32}\;$y implies $\tau\ge 10^8\;$s
\cite{Batty},
comparable to the latest result with free neutrons at Grenoble  
\cite{ILL}.
The perspective of checking directly $\tau\ge10^{11}\;$s in the  
foreseen
experiment at Oak Ridge corresponds to limits on $T$ which are not
conceivably reachable in underground experiments.

\section*{Acknowledgements}
J.M.R.\ would like to thank the organisers for their kind invitation,  
making
possible many stimulating and fruitful discussions with other  
participants.
The hospitality provided by the theory group at Oak Ridge, and in
particular  T.D. Barnes, is also gratefully acknowledged.

\end{document}